# Direct comparison of ARPES, STM, and quantum oscillation data for band structure determination in $Sr_2RhO_4$

I. Battisti[1], W.O. Tromp[1], S. Ricco[2], R.S. Perry[3,4], A.P. Mackenzie[5,6], A. Tamai[2], F. Baumberger[2,7], M.P. Allan[1]

[1] *Leiden Institute of Physics, Leiden University, Niels Bohrweg 2, 2333 CA Leiden, The Netherlands*
[2] *Department of Quantum Matter Physics, University of Geneva, 24 Quai Ernest-Ansermet, 1211 Geneva 4, Switzerland*
[3] *London Centre for Nanotechnology and UCL Centre for Materials Discovery, University College London, London WC1E 6BT, United Kingdom*
[4] *ISIS Facility, Rutherford Appleton Laboratory, STFC, Chilton, Didcot, OX11 0QX, United Kingdom*
[5] *Max Planck Institute for Chemical Physics of Solids, D-01187 Dresden, Germany*
[6] *Scottish Universities Physics Alliance, School of Physics and Astronomy, University of St. Andrews, St. Andrews, Fife KY16 9SS, United Kingdom*
[7] *Swiss Light Source, Paul Scherrer Institute, CH-5232 Villigen PSI, Switzerland*

Email: Robin.Perry@ucl.ac.uk, Andy.Mackenzie@cpfs.mpg.de, Felix.Baumberger@unige.ch, Allan@physics.leidenuniv.nl

**ABSTRACT**

Discrepancies in the low-energy quasiparticle dispersion extracted from angle resolved photoemission, scanning tunneling spectroscopy and quantum oscillation data are common and have long haunted the field of quantum matter physics. Here, we directly test the consistency of results from these three techniques by comparing data from the correlated metal $Sr_2RhO_4$. Using established schemes for the interpretation of the experimental data, we find good agreement for the Fermi surface topography and carrier effective masses. Hence, the apparent absence of such an agreement in other quantum materials, including the cuprates, suggests that the electronic states in these materials are of different, non-Fermi liquid like nature. Finally, we discuss the potential and challenges in extracting carrier lifetimes from photoemission and quasiparticle interference data.

**INTRODUCTION**

Strongly correlated electrons are at the root of some of the most mysterious quantum materials, including unconventional superconductors, strange metals, and heavy fermion materials[1–5]. Most of the exotic phases of electronic matter in these systems emerge from collective behavior of the electrons. A universally accepted understanding of these systems is still lacking, and requires close cooperation between scientists using different theoretical and experimental methods. From the experimental side, many insights to date have come from spectroscopic techniques that probe the band structure and many-body renormalizations of electrons close to the Fermi level, including angle-resolved photoemission (ARPES), scanning tunneling microscopy (STM), and quantum oscillations (QO), which are the focus of this article.

In the most widely used interpretations, spectroscopic-imaging STM (SI-STM) and ARPES probe the spectral function in real and reciprocal space[6–10]. Quantum oscillations probe the Fermi surface area and the $k$-averaged cyclotron mass which can in turn be related to the pole of the spectral function at energies close to the Fermi level[11]. There should thus be well-defined relations between the quantities measured by these three techniques[12].

Surprisingly though, several apparent contradictions between results based on these techniques can be found in the literature. Such contradictions can involve very fundamental properties of the electronic structure: for example, quantum oscillation studies on underdoped cuprate high-temperature superconductors claim the existence of Fermi surface pockets while STM and ARPES reported disconnected Fermi arcs[13,14]. Similarly, the strength of gap inhomogeneities seen by STM in several unconventional superconductors appears to be inconsistent with gap broadening in ARPES spectra that average over large areas. These and other discrepancies between results of different techniques have previously been discussed in cuprate superconductors[15–18], heavy fermion systems[19] and topological insulators[20]. However, it often remains unknown if these apparent differences are a consequence of some inherent limitations of the techniques or if they are due to challenges of data interpretation, also connected to the exotic non-Fermi-liquid nature of some of these systems. Given this lack of understanding, discrepancies are frequently attributed to the use of samples grown in different research laboratories or are ignored because of a lack of trust in one of the techniques.

With this article, we aim to test the consistency of data from ARPES, STM and QO experiments by making an unbiased comparison on the same correlated electron material. The ideal candidate for such a comparison should be a quasi-two-dimensional (2D) metal in which electron correlations still play an important role, but without the mysteries associated with materials like unconventional superconductors. Ideally it should further be structurally similar to the cuprates, ruthenates and iridates. Such a material could then act as a representative for the wider class of transition metal oxides, but -in contrast to cuprates, ruthenates and iridates- is well understood and simple enough that it can clearly be described within Fermi liquid theory. With this in mind, we chose $Sr_2RhO_4$, a layered perovskite that fulfills the conditions above.

**RESULTS & DISCUSSIONS**

*ARPES and QPI Fermi Surface*

In Figs. 1 and 2, we compare Fermi surface data from the three techniques. Consistent with previous reports[11,21,22], the ARPES $k$-space map (Fig. 1a) shows two nearly circular contours that are backfolded to form 3 pockets; a hole-like α pocket centered at Γ, a lens-shaped electron pocket at M ($β_M$) and a square-shaped hole pocket at X ($β_X$). The backfolding is of structural origin and arises from a staggered rotation of the $RhO_6$ octahedra around the c-axis, which doubles the in-plane unit cell. Hybridization with $e_g$ states pushes the *xy* band of $Sr_2RhO_4$ below the chemical potential, leaving a Fermi surface with out-of-plane *xz/yz* character, containing 3 electrons per Rh site[11,22,23]. Despite the quasi-1D hopping associated with the out-of-plane orbitals, the Fermi surface is nearly isotropic. This change arises from a strong level repulsion of states that would be degenerate in the absence of spin-orbit coupling[24,25]. The marked anticrossing can be attributed to an enhancement of spin-orbit splittings in the presence of electronic correlations[25–27].

Figure 1b,c show an STM topography and a constant energy conductance layer, where spatial modulations attributed to quasiparticle interference are neatly resolved. The few atomic defects

in the field of view clearly act as scattering centers for quasiparticles, creating the interfering standing wave patterns. The Fourier-transform of the normalized conductance layer at the Fermi level E=0 meV is shown in Fig. 1d. To mitigate the set-up effect, we take the Fourier transform not of the conductance layers $dI/dV(\mathbf{r},eV)$, but of the normalized conductance data, $dI/dV(\mathbf{r},eV)/(I(\mathbf{r},eV)/V)$, where $I(\mathbf{r},eV)$ is the tunnelling current and $V$ is the bias voltage (see discussion in the methods and Supplementary Figure 11)[28–32]. For the β band, we directly observe the STM 'Fermi surface' with wave-vector $\mathbf{q}=2\mathbf{k}_F$. More generally, we expect to observe features corresponding to scattering vectors $q$ that connect points of high spectral weight in momentum space. For the present case, we can readily connect these $q$ vectors with the Fermi surface measurement from ARPES. Interestingly, different scattering processes have different strength. While some scattering processes are very clear, others are less visible or completely absent. Varying intensities or absences of scattering processes have been observed in other materials[33-35] and can stem from the differences in the scattering process. For example, different QPI scattering intensities are expected from magnetic versus potential scattering or from broad coulombic potentials versus localized impurity potentials[36-40]. In principle, theoretical tools exist to predict QPI intensities based on both the electronic structure of the material, and the nature of the scattering potenial[36-40]. A comparison with such QPI simulations could allow us to learn more about the defect states in $Sr_2RhO_4$.

In Fig. 1e, we use the Fourier transform of the SI-STM data discussed above to reconstruct the entire FS of $Sr_2RhO_4$ from the QPI pattern. To this end, we first extract peaks in the data by fitting the intensity profiles in radial cuts (see Supplementary Figure 2). We then obtain the fundamental β band in $k$-space by rescaling the $q$-vectors by a factor of two. The backfolded β bands are obtained by translating the fundamental band by reciprocal lattice vectors determined from the STM topography. The α band is reconstructed by subtracting the interband vector $\mathbf{q}_{\alpha-\beta}$ from the intraband scattering $\mathbf{q}_{\beta-\beta}$. In Fig. 1e, we display in blue the data points derived in this way, and in grey the direct interband scattering vectors $\mathbf{q}_{\alpha-\beta}$ that we used for the derivation. The identification of the grey scattering vectors as $\mathbf{q}_{\alpha-\beta}$ is corroborated by the Fermi velocity we obtain for this vector (see below). The direct comparison of these contours with the ARPES Fermi surface shows good agreement for all Fermi surface pockets, as further illustrated in Fig. S3.

*Shubnikov de Haas Oscillations*

Next, we use QO to find the Fermi surface areas and quasiparticle masses. The inset of Fig. 2a shows a trace of quantum oscillations in the magnetoresistance at 0.1 K (Shubnikov de Haas (SdH) oscillations). To extract Fermi surface information from the quantum oscillations measurement, we analyse the frequency components plotted in Fig. 2a. Seven closely spaced peaks are resolved, corresponding to seven frequencies between 0.9 kT and 1.3 kT. This might be surprising at first, as from our previous analysis we expect only three distinct Fermi surface pockets. We attribute the higher number of QO frequencies to two effects. Firstly, the finite interlayer hopping implies that the Fermi surface of $Sr_2RhO_4$ is quasi-cylindrical, and thus has multiple extremal orbits per sheet. The characteristic signature of such a remnant 3D Fermi surface warping is an overall $1/\cos(\theta)$ field angle dependence of the frequencies (consistent with quasi-2D electronic structure) with small splittings that disappear for certain angles, as observed in Fig. 2b. The quasi-2D nature of the quasiparticle band structure is confirmed directly by photon energy dependent ARPES measurements (Fig. 2c) probing the Fermi surface along $k_z$. Secondly, the ARPES measurements resolve a small splitting in the β-band along ΓM. This small degeneracy lifting can be attributed to the doubling of the unit cell along the c-axis and is reproduced by LDA+U+SO band structure calculations[25]. Hence, there are four primary frequencies up to the measured out-of-plane angle of 40°. We can then use multiple facts to constrain the band assignments: (i), Following ARPES and STM data, the extremal orbit areas increase in size from the α (hole), $β_M$ (electron) and $β_X$ (hole). (ii), The total electron count should be 3 electrons per Rh atom. (iii), The experimental specific heat $γ$ can be calculated in the 2D approximation from:

$$γ = (πN_A k_B^2 a^2)(3ℏ^2)^{-1}(m_α + 2m_{βM} + m_{βX}) \quad (1)$$

where $N_A$ is Avogadro's number, $k_B$ is Boltzmann's constant, $a$ is the tetragonal lattice parameter (3.857 Å) and $ℏ$ is Planck's constant. (iv), Following the ARPES data, the $β_M$ band should be split leading to two frequencies.

Combining these conditions, we draw the conclusion that α corresponds to the lowest frequency (0.93 kT, corresponding to 1.934 electron/Rh, see methods), $β_M$ to the two middle frequencies (average 1.068 kT, 0.152 electrons/Rh) and $β_X$ to the highest frequency (1.288 kT, 0.908 electrons/Rh). The calculated $γ = 17.4 ± 0.8$ mJ/Rh mol K$^2$ then agrees with the directly observed value of 17.7 ± 0.7 mJ per Rh mol K$^2$ [22], and the total electron count is 2.994 electrons per Rh. A quantitative comparison of the Fermi surface volumes extracted from ARPES, QO and STM based on this assignment can be found in Table 1. QO amplitudes also yield the Dingle temperatures (1.5K for the α pocket, 1.5K for the $β_M$ pocket, and 1.9K for the $β_X$ pocket), which can be related to the mean free path of the electrons. We refer to the literature for a detailed discussion on the challenges of such an interpretation[41].

*Quasi-particle dispersion*

We now turn our attention to the low-energy dispersion. In Fig. 3, we show constant energy layers for selected energy levels and the energy-momentum dispersion along the two high symmetry directions for both ARPES and SI-STM. These data confirm that $β_M$ is an electron pocket while α and $β_X$ are hole-like. The ARPES data also reproduce the splitting of the β band along ΓM observed in Fig. 2. The STM dispersions plots show several features that are not observed by ARPES. These can all be assigned to different β-β intraband and α-β interband scattering vectors translated by reciprocal lattice vectors. The **q** vector which is most clearly resolved by STM along both high-symmetry directions arises from β-β intraband scattering. Comparing its dispersion with the β-band measured by ARPES, we find quantitative agreement along ΓX, where both techniques lead to measured Fermi velocities $v_F = 0.55$eVÅ. Along ΓM, where band structure calculations find a small splitting in the Fermi surface, our ARPES data resolves both bands and shows that they have slightly different dispersion with Fermi velocities of 0.57eVÅ and 0.77eVÅ, respectively, compared to $v_F = 0.70$eVÅ extracted from the STM dispersion. The lack of a noticeable splitting in the STM dispersion cannot be explained by insufficient momentum resolution, suggesting that it is due to a vanishing STM matrix element for one of the bands.

In order to extend this comparison to QO, we look at the quasiparticle cyclotron masses *m\**. These masses can be deduced for individual Fermi surface pockets from the temperature dependence of the quantum oscillation amplitudes using the Lifshitz-Kosevich formula (Supplementary Figure 4). For a 2D Fermi surface, they can also be calculated without any approximations from the full mapping of the low-energy quasiparticle band structure obtained by STM and ARPES, using $m^* = \frac{\hbar^2}{2\pi}\frac{dA_{FS}}{dE}$ where $A_{FS}$ is the Fermi surface volume. To this end, we extract the areas of the pockets not only at the Fermi energy, but at a few constant energy layers within a small window. The linear fits of these areas shown in Fig. 4 yield the effective masses of the different pockets. We note that the slope $dA_{FS}/dE$ decreases strongly near the chemical potential in the ARPES data while no such effect is observed in STM. This change of slope is a known artifact arising from the combination of a Fermi cutoff and finite energy resolution. For the quantification of *m\**, we thus exclude a narrow energy range around $E_F$ from the ARPES data. Table 1 shows the values of the effective masses obtained by STM, ARPES, and QO measurements. Knowledge of *m\** and the Fermi surface area also allows a sheet-averaged Fermi velocity to be calculated from QO data as $v_F = \hbar(A/\pi)^{1/2}/m^*$. These values are also shown in table 1 for comparison with STM and ARPES.

*Lifetime analysis*

Finally, we discuss the extraction of peak widths in the ARPES and STM data, which can in principle be related to quasiparticle lifetimes. Here, the two techniques face rather different challenges. In simple systems, like $Sr_2RhO_4$ studied here, the measured photoemission intensity appears to represent the spectral function. For a sufficiently linear band, the imaginary part of the self-energy (which is equal to half the inverse lifetime in a Fermi liquid) can thus be obtained from the width $W_k$ of the momentum distribution curves (MDC) as $\Sigma''(k, \omega) = W_k(\omega) * v(\omega)/2$, where $v(\omega)$ is the slope of the dispersion at the same energy. In practice, the main difficulty is the treatment of the effective resolution of ARPES measurement. ARPES peak widths measured at state-of-the-art instruments are rarely limited by the instrumental energy and momentum resolution but contain a variety of other contributions that are notoriously hard to quantify. These include in particular broadening from the finite integration over perpendicular momenta, structural mosaicity in the probed area and the often-unknown quality of the surface. Additional broadening can occur from work function inhomogeneities of and around the sample which cause uncontrolled electric fields that degrade the resolution of the electron optics. Finally, in an energy range of ~ *dE*/2, where *dE* is the effective energy resolution, the MDC peak position starts to deviate noticeably from the intrinsic quasiparticle pole, which prohibits a model-free analysis of very low energy dispersions and self-energies. STM does not suffer from these experimental difficulties. However, it is not always clear to what extent the tunneling spectra reflect $A(\mathbf{r}, \omega)$. In particular the so-called setup effect, the dependence of the tunneling spectra on the lateral variation of the tip-sample distance, which itself is defined by the setup current and voltage, can cause complications. In many cases, the set-up effect can be mitigated by taking ratios between different quantities as we do here, but then the interpretation of the resulting data is less straight forward. In addition, analyzing self-energies from STM data can be complicated when different **q** vectors overlap, especially for complex Fermi surfaces. We also note that a unique reconstruction of the spectral function from STM data is not always possible. Perhaps most importantly, one has to consider the scattering mechanism, which can strongly influence line-shapes and line-widths[36-40]. For this reason, only few attempts have been made to extract lifetimes from STM data[42-45].

Despite these difficulties, an analysis of the MDC's along the ΓX direction shown in Fig. 5a-c clearly show an energy dependence of the quasiparticle lifetime in fair agreement with the expectations for a Fermi liquid-like metal. In a Fermi liquid, we expect the imaginary part of the self-energy to be a quadratic function at low energy, $\Sigma''(\omega) = A\omega^2$, where $A$ is a material specific prefactor. To compare our results with the expectation, we plot the linewidths and self-energies extracted from ARPES and STM data as a function of energy in Fig. 5d,e (for a comparison of the individual ARPES and STM spectral lines, see Supplementary Figure 6). Indeed, our results are consistent with a quadratic dependence on energy. Further, the absolute scale of the measured self-energy is of the same order than what is obtained for $Sr_2RhO_4$ with dynamical mean field theory (DMFT)[50,51] calculated for a generic two-dimensional Fermi liquid using the random phase approximation[46-49]. We note that the agreement holds for an energy range of that is a significant fraction of the Fermi energy, which is roughly 400meV for the $\beta$ band when defined in the parabolic band approximation as $E_F = \hbar^2 A_{FS}/2\pi m$. This is encouraging for further lifetime investigations: given a good understanding of both the resolution for ARPES and the scattering process for STM, both techniques have the potential to bring insight into energy and momentum dependent correlation effects in electronic matter of quantum materials.

We show SI-STM results of $Sr_2RhO_4$ and extract its Fermi surface and low-energy dispersion, and present a quantitative comparison of the STM quasiparticle interference data with ARPES and QOs. Our data here reveals the previously unknown band structure above the Fermi level, and quasiparticle lifetimes for $Sr_2RhO_4$. We show that Fermi surface volumes agree among the three techniques within ~ 1% of an electron for all pockets, while quasiparticle masses exhibit a relative variation of ~ 30%. We consider these values to be characteristic for the precision that can realistically be obtained in favorable cases with these three techniques, and therefore conclude that for the oxide $Sr_2RhO_4$, STM, ARPES, and QO can extract the same information regarding Fermi surface and low energy dispersion. The relevance of our study goes beyond $Sr_2RhO_4$: Our data suggest that apparent disagreements in the literature on cuprates do not arise from the intrinsic structural complexity of oxides but are likely a consequence of our limited understanding of materials with non-Fermi liquid electronic states and the applications of the techniques to such samples, especially ones with significant spatial inhomogeneity.

**Methods**

*Sample preparation*

Our single crystals samples were grown in a *Crystal Systems* four mirror image furnace using a flux feeding floating zone method. Dried $SrCO_3$ and $Rh_2O_3$ (3N) were ground together in a 1:0.575 ratio, pelletised and calcined at 1000°C in flowing $O_2$ atmosphere for 24 hours. Rods were hydrostatically pressed using the usual methods and sintered at 1100°C for 2 hours in flowing $O_2$. The growth conditions in the image furnace were 100% $O_2$ gas at 10 bar pressure, growth speed of 10 mmhr$^{-1}$ and a counter rotation of 30 rpm. Subsequently, the crystals were annealed 1150°C under flowing oxygen for 2 weeks, as described elsewhere.[22]

The surfaces studied by STM and ARPES have been obtained by cleavage in ultrahigh vacuum.

*Quantum oscillations:*

Quantum oscillations measure low energy characteristics of the electron fluid in an applied magnetic field. The oscillations, caused by the Landau quantization from the magnetic field, give precise information on the size of the Fermi pockets and the effective masses of the electrons. Quantum oscillations are a true bulk probe that is generally not influenced by surface effects but they are very sensitive to disorder in the crystals and require high quality samples to be observed. Furthermore, they also require high magnetic field and low temperatures to suppress the quasiparticle-quasiparticle scattering and the interpretation is not always simple as little information is given about the loci, shape and type (electron or hole) of the Fermi pockets. When a strong magnetic field **B** is applied to the sample, the Landau quantization of quasiparticle orbits leads to an oscillation of the density of states at the Fermi level, periodic in reciprocal field. These oscillations are reflected in most of the physical properties; in the case of magnetoresistance they are called Shubnikov–de Haas (SdH) oscillations[41,52]. By analyzing the frequency $f$ (in Tesla) of the oscillations across an inverse field range, the number and sizes of the Fermi surface pockets can be obtained. Moreover, the effective masses for the various pockets can be deduced from the temperature dependence of the oscillation amplitude (Fig. S4) via the Lifshitz-Kosevich formula, although, we note that the data analysis can be non-standard when measuring across a broad magnetic field range (for a comprehensive discussion see Ref. 22). The QO amplitudes also contain the Dingle temperature. These can be used to find mean free paths of 500 Å for the α pocket, 714 Å for the $β_M$ pocket, and 481 Å for the $β_X$ pocket.

Quantum oscillation data was acquired using a standard four probe technique in a dilution refrigerator (current $I$ = 300 μA) for temperatures between 0.1 K and 1.0 K and magnetic fields between 7 T and 15 T. Low contact-resistance electrical connections were made to the crystals using gold wire (25 micron) and *Dupont 6838* high temperature curing paint (annealed at 470°C under O2). The current was applied in the ab plane (the two-dimensional morphology of the crystals allowed for easy identification of the crystallographic ab plane and c axis). In the dilution refrigerator the samples were mounted on an in-situ single axis rotator for the angular quantum oscillation study. Three crystals were measured from the same batch, with consistent results.

*ARPES:*

ARPES measures single particle excitations directly in momentum space. The most commonly used expression for the photocurrent $I(\mathbf{k}, \omega)$ is:

$$I_{ARPES} = I_0 \left|M_{f,i}\right|^2 f(\omega)\, A(\mathbf{k}, \omega) * R(\delta\mathbf{k}, \delta\omega) \quad (2),$$

where $M_{f,i}$ represents the photoemission matrix elements, $A(\mathbf{k}, \omega)$ is the spectral function and $f(\omega)$ the Fermi function[10]. The expression for the intrinsic photocurrent is then convolved with the experimental momentum and energy resolution $R(\delta\mathbf{k}, \delta\omega)$. Besides experimental difficulties, complications can arise from the interference of photocurrents from different emission sites and/or from different terms in the light-matter interaction Hamiltonian. Expressing the photocurrent in terms of the spectral function further relies on the sudden approximation, i.e. the assumption that the photoexcitation is instantaneous and that there is no interaction between photoelectron and the sample during the photoemission process[10]. This approximation is well tested down to much lower photon energies than used in the present work.

The ARPES experiments reported in this paper have been performed at beamline I05 of Diamond Light source using photon energies in the range of 20 – 80 eV[53]. Energy and momentum resolutions were set to ~ 5 meV / 0.008 Å$^{-1}$, except for the data shown in Fig. 2c where the resolution varied with photon energy and thus with $k_z$. All data were acquired at T ~ 8K.

*STM:*

STM measures the tunneling current generated between an atomically sharp tip and a conducting sample when a voltage $V$ is applied between the two. By scanning the tip over the sample surface, STM directly delivers real-space information with atomic resolution[54]. The tunneling current $I$ is directly proportional to the *integrated local* density of states (LDOS) of quasiparticles, which in the formalism of many-body physics can be defined via the *local* spectral function $A(\omega, \boldsymbol{r}) = \sum_k A(\boldsymbol{k}, \omega; \boldsymbol{r})$. The local spectral function of the sample can be accessed for both occupied and unoccupied states by measuring the local differential conductance[52]:

$$\frac{dI}{dV}(eV, \mathbf{r}) = \frac{4\pi e^2}{\hbar} |t(\mathbf{r})|^2\, A_\text{T} A_\text{S}(\omega = eV, \mathbf{r}) \quad (3),$$

where $A_{\text{S,T}}$ are the spectral functions of sample and tip, respectively, and where we approximated the Fermi-Dirac distribution as a step function. $|t(\mathbf{r})|^2$ represents the position-dependent tunneling matrix element that contains the exponential dependence on tip-sample distance. Usually, the spectral function of the tip, $A_\text{T}$ is designed to be constant and the momentum-dependence of the tunneling matrix elements is ignored.

When measuring in *spectroscopic-imaging* mode (SI-STM), for each pixel on a chosen field of view a $dI/dV$ spectrum is acquired at the tip-sample distance determined locally by the set-up conditions $(V_\text{S}, I_\text{S})$. The result of such a measurement is a three-dimensional dataset representing the local density of states as function of position and energy.

Because we determine the tip-sample distance at each point by the set-up conditions, the effect of the matrix element (assuming it is energy independent) is cancelled. However, the procedure does bring in an extra denominator: $\frac{dI}{dV}(eV,\mathbf{r}) = \frac{I_S A_S(eV,\mathbf{r})}{\int_0^{eV_S} A(E,\mathbf{r}) dE}$. The procedure can thus introduce additional artifacts into the measured differential conductance $dI/dV$[28,55], the so-called *set-up effect*.

A common way to reduce this effect is to choose set-up conditions far away from the Fermi level such that inhomogeneities in the integrated density of states average out, however, this is not always experimentally possible. Other methods include the use of the ratio between quantities with positive and negative bias[56], or the division of the differential conductance by the total conductance $(dI/dV)/(I/V)$[28–30] - the approach that we also use in this paper. See Figs. 5b-c, S1, S6 for comparisons. The current $I$ is the *measured* current at that particular location and bias $V$, which means it is small but generally non-zero at the Fermi level. The voltage $V$ is a value set in the experiment, implying for the Fermi level that the data would be multiplied by 0 in the normalization. To circumvent this, we manually add a 10uV (negligible to the energy scale set by temperature) offset in data processing.

The STM experiments reported in this paper have been performed with an ultra-high vacuum, home-built STM with exceptional stability, described elsewhere[57]. All data was taken at a base temperature of 4.2K. Measurements are performed with a chemically etched tungsten tip that is prepared by field emission on a gold surface before measuring $Sr_2RhO_4$.

**Acknowledgements**

This work was supported by the UK-EPSRC under grant EP/G007357/1, by the Swiss National Science Foundation (SNSF) under grants 200020_165791 and 200020_184998, by the Max Planck Society, by the European Research Council (ERC StG SpinMelt) and by the Netherlands Organization for Scientific Research (NWO) under grants 680-47-536 and FOM-167. We acknowledge Diamond Light Source for time on beamline I05 under proposals no. SI13398 and SI5282

**Figures**

|  |  | α | β$_M$ | β$_X$ |
|---|---|---|---|---|
| $v_F$ (eV Å) | QO | 0.41* | 0.47* | 0.49* |
|  | ARPES | 0.41 | 0.57/0.77 | 0.55 |
|  | STM | 0.41* | 0.70 | 0.55 |
| $<A>$ (%) | ARPES | 7.7% | 7.9% | 9.1% |
|  | STM | 6.8% | 7.1% | 7.9% |
|  | QO <peaks> | 6.6% | 7.6% | 9.2% |
|  | QO single peak | 6.6% | 7.1% | 9.1% |
|  |  |  | 7.3% | 9.3% |
|  |  |  | 7.9% |  |
|  |  |  | 8.2% |  |
| $<m>$ ($m_e$) | QO | 3.1 | 2.9 | 3.1 |
|  | ARPES | 3.3 | 2.2 | 3.3 |
|  | STM | - | 2.4 | 2.9 |

**Table 1 Summary of band structure parameters from different techniques.** Comparison between values obtained from the three techniques. $v_F$ is the Fermi velocity, $A$ (in % of the reduced tetragonal Brillouin zone) and $<m>$ (in units of $m_e$) are the Brillouin zone filling and average mass for each of the three sheets, respectively. The pocket-averaged Fermi velocities from QO (marked with *) are extracted using $\hbar k_F = m v_F$, using the pocket-averaged Fermi wave vector. The Fermi velocity for the $\alpha$ band from STM (also marked with *) was extracted from the slope of the $\mathbf{q}_{\alpha\beta}$ and $\mathbf{q}_{\beta\beta}$ signals. The QO values for $A$ are given both as average over the multiple peaks for each pocket, and for every peak.

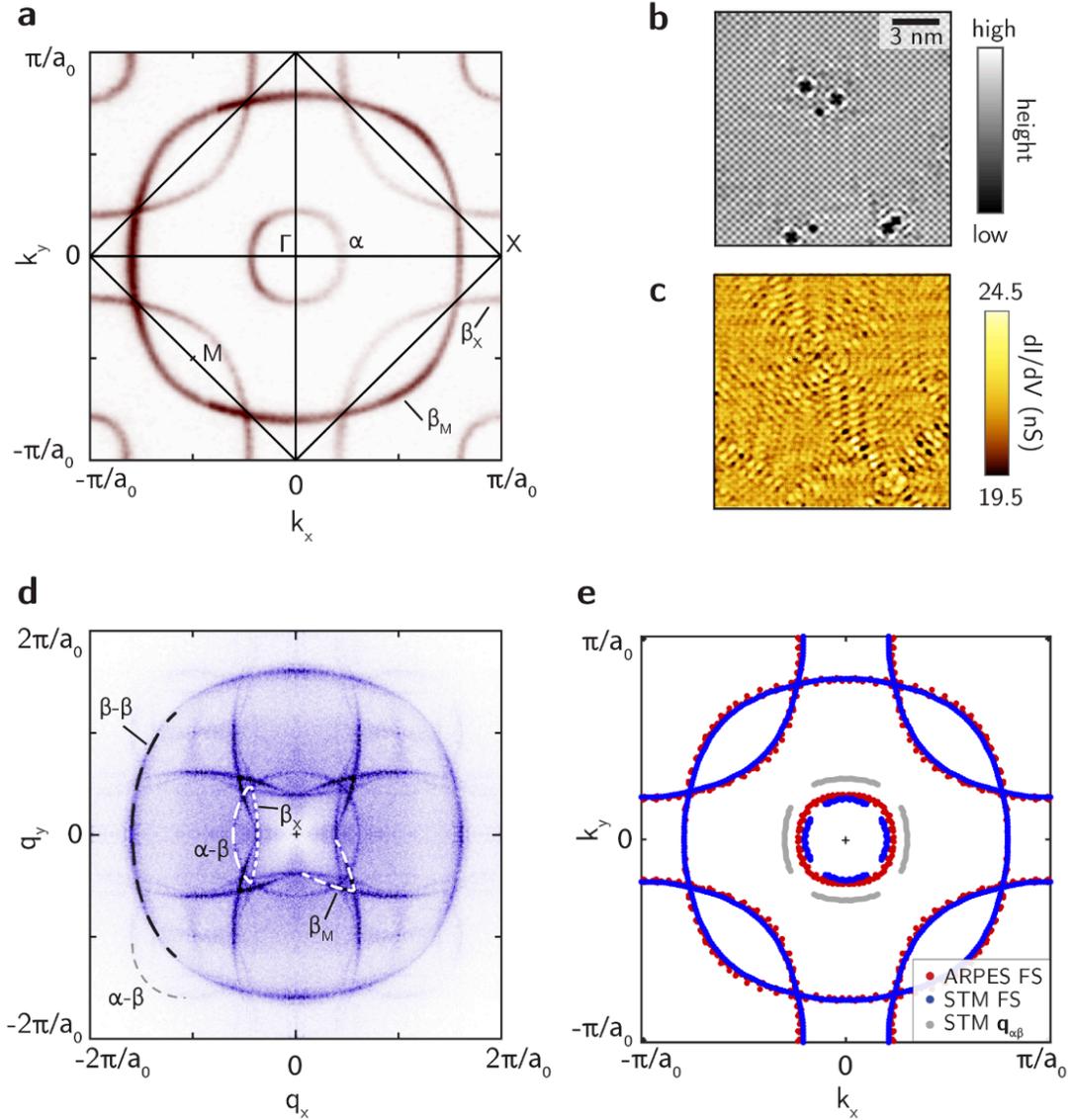

**Figure 1 Sr$_2$RhO$_4$ Fermi surfaces. a**, ARPES Fermi surface. **b**, STM topograph with atomic resolution showing presence of impurities. **c**, STM conductance layer $dI/dV(\mathbf{r},eV)$ at energy $eV$=-20meV, acquired simultaneously to the topograph in panel b, showing interference between quasiparticles standing waves patterns. **d**, STM 'Fermi surface', obtained by a two-dimensional Fourier transform of the conductance layer corresponding to the Fermi level of a spectroscopic map measured over a field of view of 70×70nm$^2$. Here and for all QPI data, we show the normalized data, i.e. the Fourier transform of $dI/dV(\mathbf{r},eV)/(I(\mathbf{r})/eV)$, to mitigate the set-up effect (see Methods). The data is additionally symmetrized and the low-q components are suppressed with a 2D gaussian: raw data is shown in Supplementary Figure1. **e**, Comparison between extracted Fermi surfaces of ARPES and STM. For completeness, we also show the QPI signal of $\mathbf{q}_{\alpha\beta}$, which was used in the derivation (see text and Supplementary Figures 2 and 3).

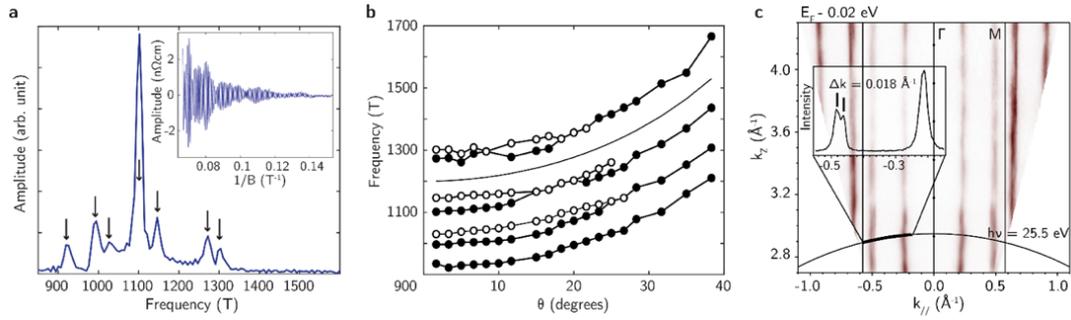

**Figure 2 Quantum oscillations and $k_z$ dispersion. a**, Shubnikov-de Haas oscillations at 0.1 K for a magnetic field parallel to the *c*-axis. The main panel shows the frequency components obtained by Fourier transform of the quantum oscillation trace shown in the inset. The peaks correspond to Fermi surface pockets. The background subtraction used was a third order polynomial and the field sweep rate was 0.05 T/min. The noise level is 50 pVHz$^{-1/2}$. **b,** Angle dependence of the QO frequencies for angle $\vartheta$ from the *c*-axis. The solid black line is a $1/\cos\vartheta$ dependence expected for a quasi-two-dimensional Fermi surface. **c**, ARPES $k_z$ dependence in the MΓ high symmetry direction at E=-20meV, showing only slight modulations of the band along the *c*-axis.

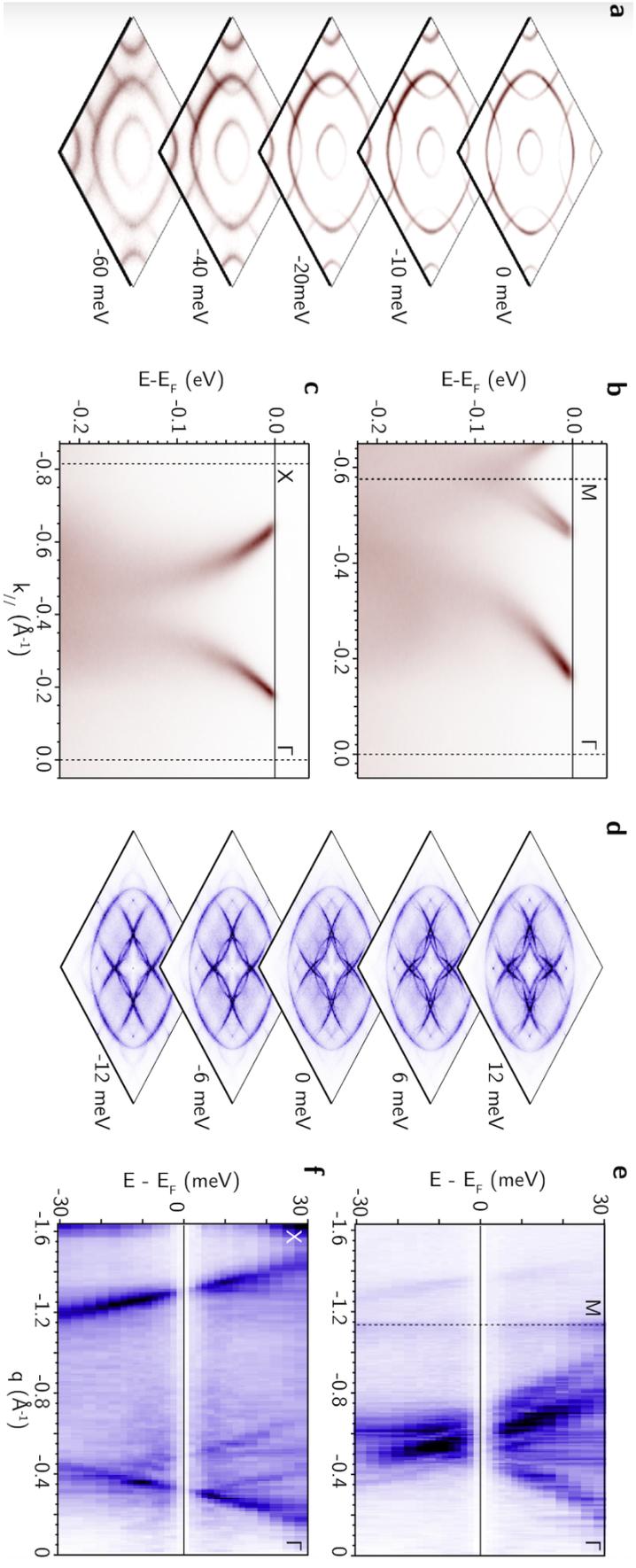

**Figure 3 ARPES and QPI dispersions along high symmetry directions. a,** ARPES constant energy layers at and below the Fermi level. **b-c,** ARPES dispersions along the high symmetry directions MΓ and XΓ. **d,** Fourier transform of normalized STM conductance layers $dI/dV(\mathbf{r},eV)/(I(\mathbf{r})/V)$ at selected energies around the Fermi level. **e-f,** Dispersions of the scattering vectors **q** obtained from STM QPI for the two high symmetry directions MΓ and XΓ, where X and M are defined as for ARPES but at double the reciprocal vectors. To improve signal-to-noise, the dispersions are obtained by averaging 10 cuts in a radial span of ±5° from the high symmetry direction. The intensity around the Fermi level is suppressed due the normalization with the total conductance $(dI/dV)/(I/V)$, as explained in the Methods.

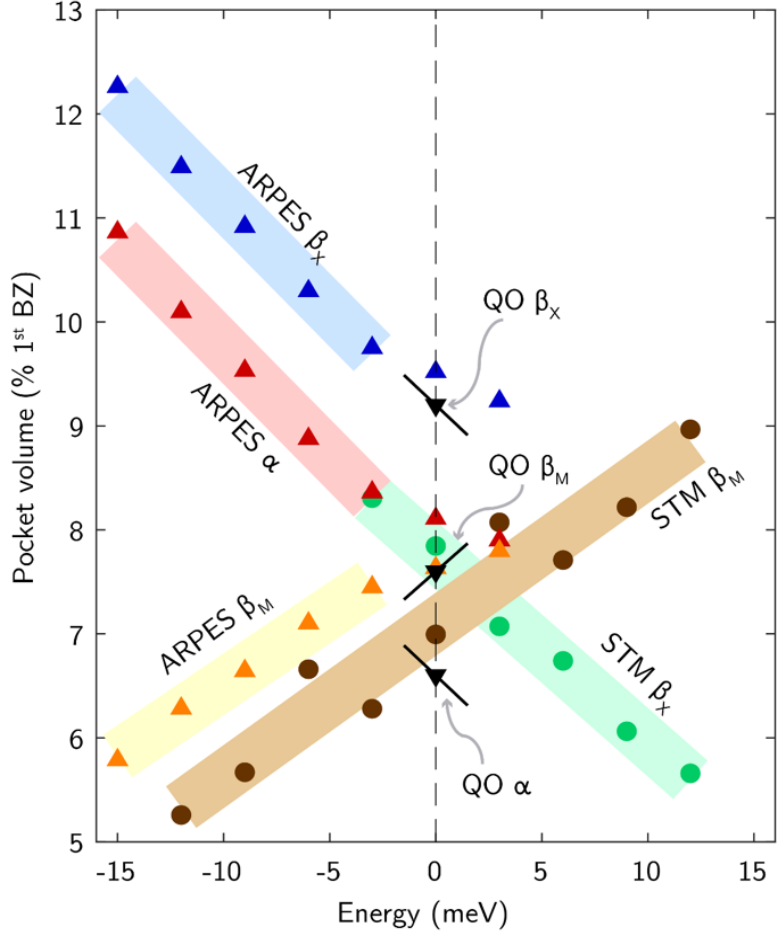

**Figure 4 Fermi surface pockets and effective masses.** Extraction of effective masses from ARPES, STM, and QO. The data points show the volume d$A_{FS}(\omega)$ of the different pockets as a function of energy (see also Supplementary Figure 5). The effective masses are proportional to the slope of d$A_{FS}(\omega)$. For the fits of the ARPES data, we excluded datapoints that are closer to the Fermi level than the energy resolution of the detector. The black lines and marks around the Fermi level indicate the masses and volumes extracted from QO.

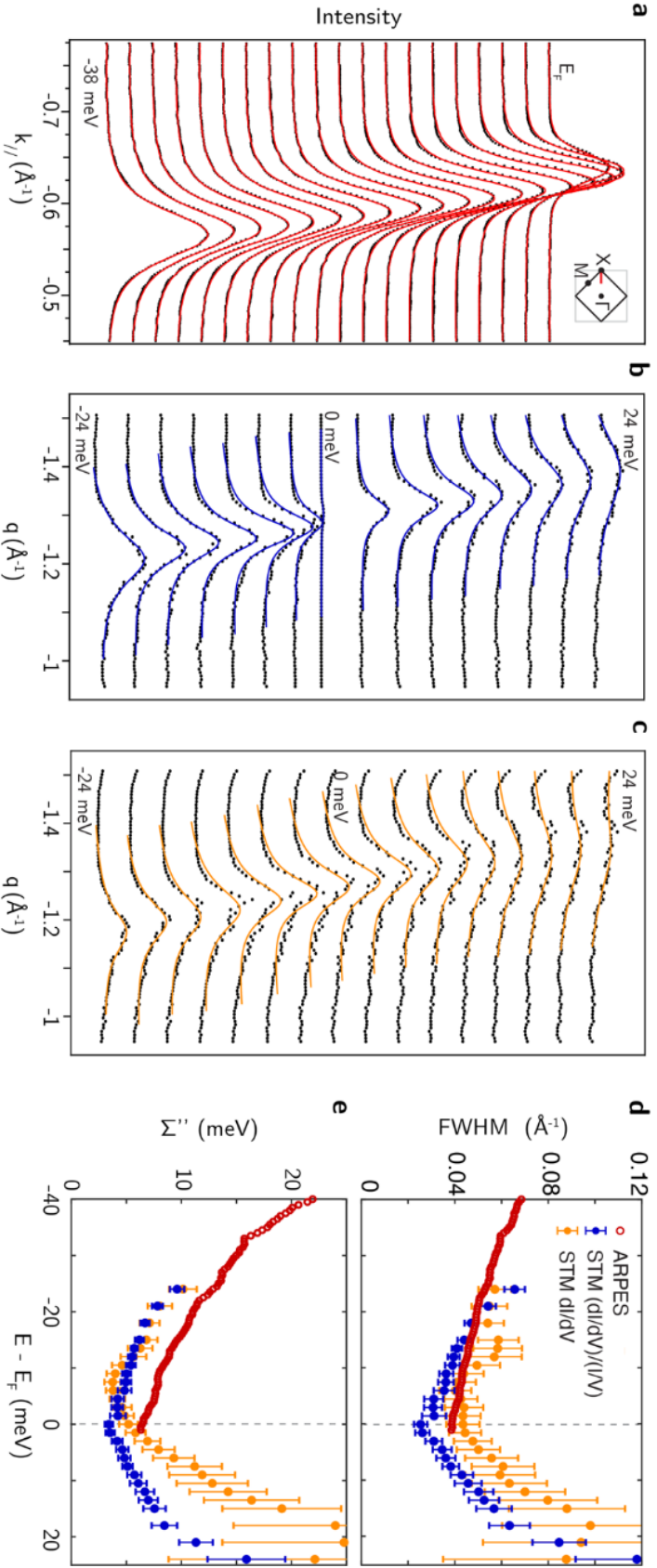

**Figure 5 ARPES and QPI linewidths and self-energies.** Linewidth analysis for the βx band. **a,** ARPES MDCs and Lorentzian fits (red). The inset shows the position of the cut as red line in the reduced Brillouin zone (black square). **b, c,** STM MDCs from normalized conductance $(dI/dV)/(I/V)$ (b) and conductance $dI/dV$ (c) including Lorentzian fits (blue and orange, respectively) with a linear background. The cuts are equivalent to the one used in panel (a) at double the reciprocal lattice vectors. **d,** Comparison of the MDC widths $W_k$ (o) from STM and ARPES data. STM widths are phenomenological full-width-half-maximum extracted by fitting a Lorentzian convoluted with a Gaussian broadening that stems from the finite resolution. **e,** Comparison of the widths multiplied by the slope of the dispersion, which, in a simplified picture, equals to the imaginary part of the self-energy. Note that the scale of the energy axis spans a significant fraction of the Fermi energy, which is roughly 400meV for the β band (in the parabolic band approximation and before hybridization, as we assume that scattering processes that are relevant for the lifetimes do not discriminate between the βx and βM bands).

**Supplementary figures:**

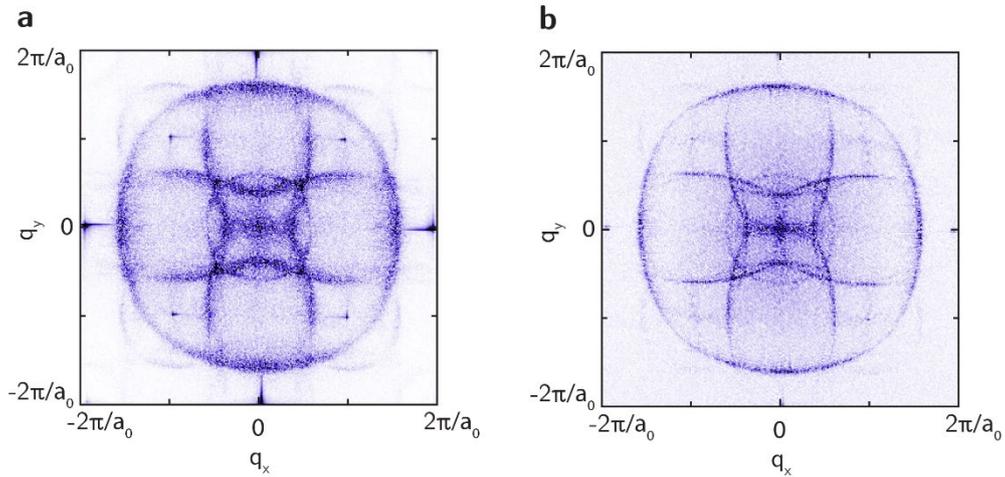

**Supplementary Figure 1.** Raw data of Fermi surfaces extracted from STM (not corrected for drift, without core suppression and not symmetrized) for both *dI/dV* (a) and (*dI/dV*)/(*I/V*) (b).

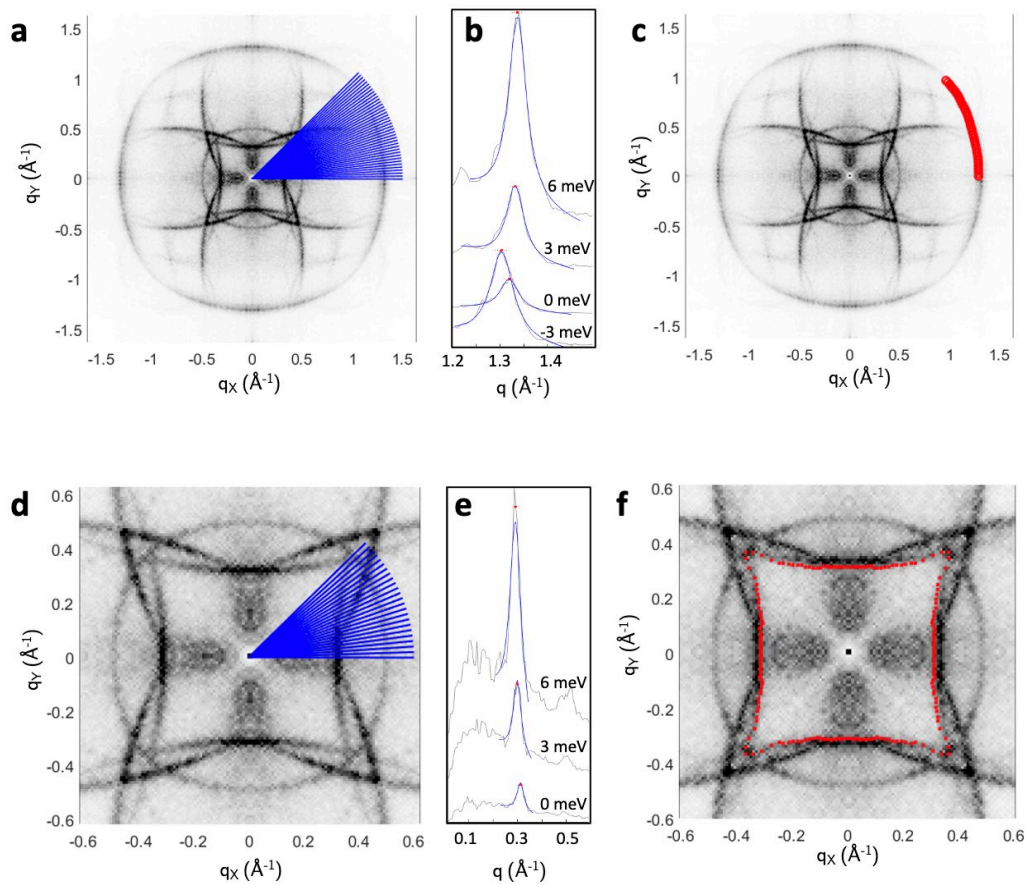

**Supplementary Figure 2.** Extraction of the points plotted in Fig. 1e and Fig. 4 for STM data. **a-c** Cuts along different angles are taken (blue lines) and fitted with Lorentzian function and a linear background. The image on the right shows the fitted points at the conductance layer at the Fermi level. **d-f** Analogous procedure for the QPI at low **q**.

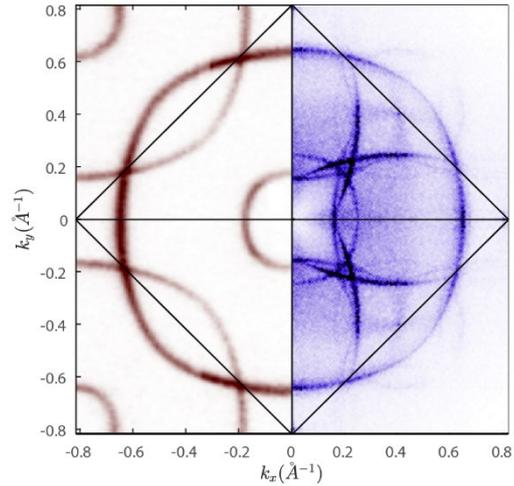

**Supplementary Figure 3.** Comparison of STM and ARPES Fermi surfaces. STM data has been processed as in Figure 1d and it is rescaled by a factor 2 to take account of the difference between scattering vectors seen with STM (*q*-space) and the direct momentum space probed by ARPES (*k*-space).

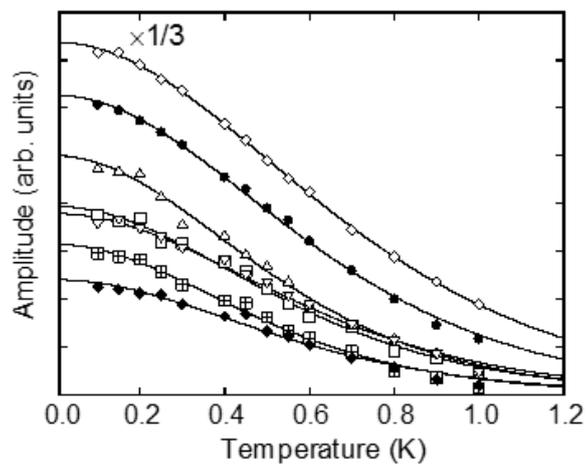

**Supplementary Figure 4.** Quantum oscillation. The dependence of the amplitude of the QO seven frequencies as a function of temperature with a Lifshitz-Kosevich fit used to extract the masses.

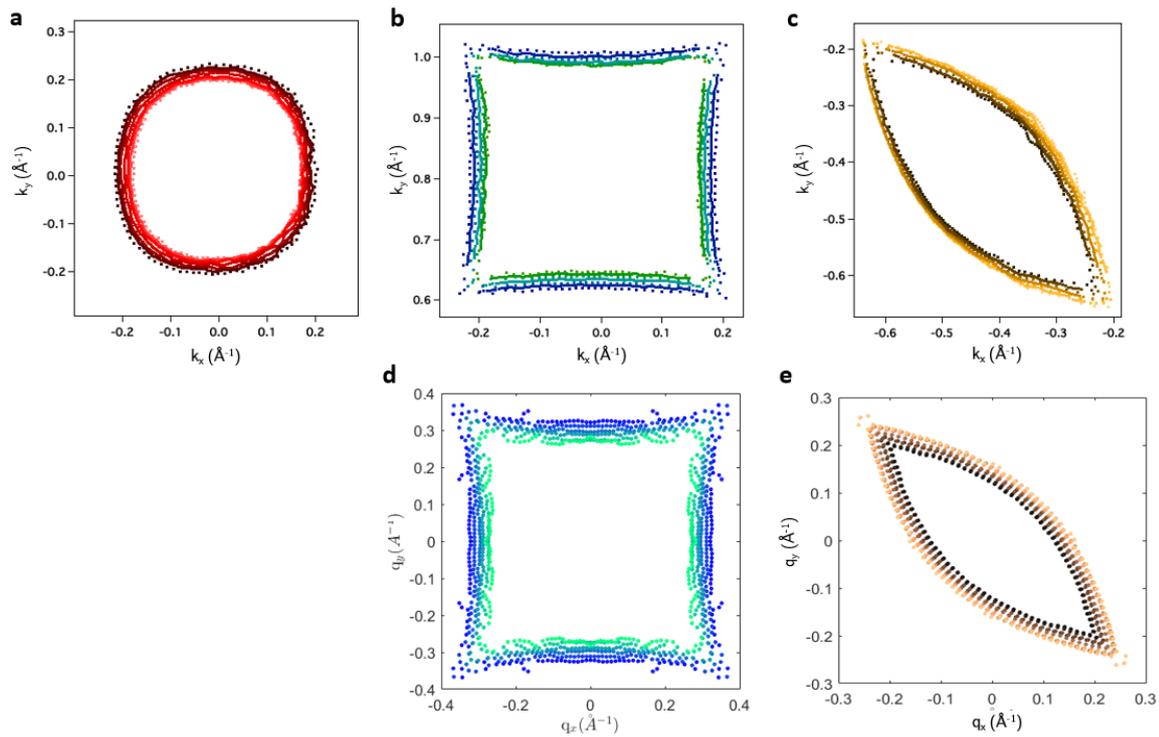

**Supplementary Figure 5.** The extracted Fermi surface pocket's volumes used for Fig. 4.

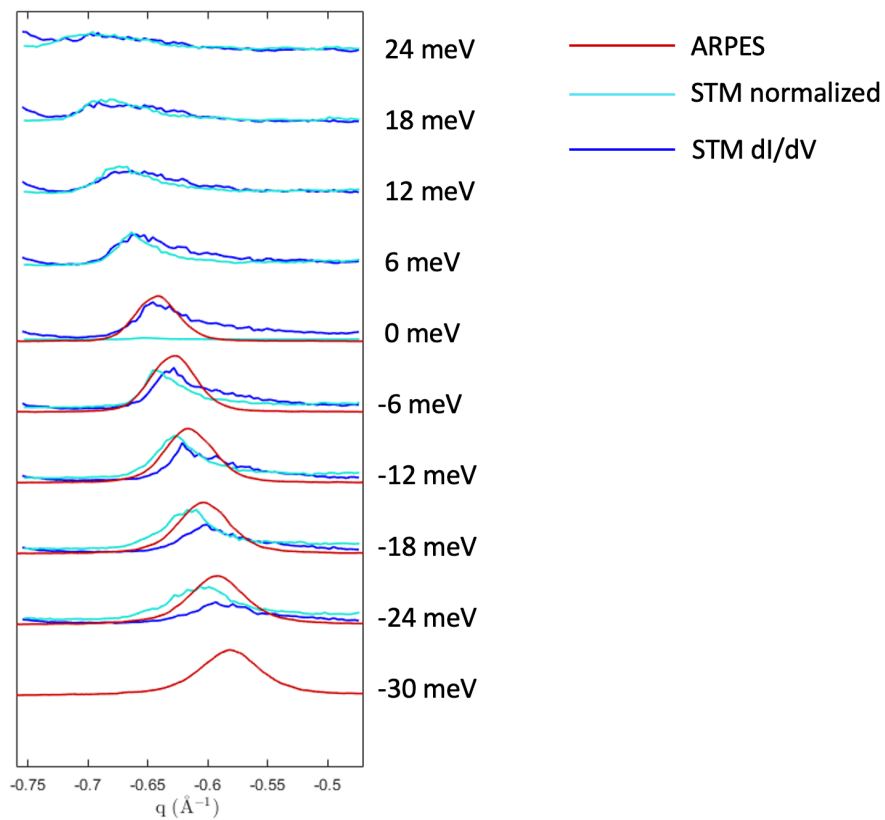

**Supplementary Figure 6.** Direct comparison of MDCs from ARPES and STM as shown in Figure 5.